\documentclass{iaus}
\usepackage{graphicx}

\title[Chemical evolution of bulges] 
{Chemical evolution of bulges at high redshift}

\author[A.Pipino, F.Matteucci, A. D'Ercole]   
{Antonio Pipino$^{1,2}$, Francesca Matteucci $^{2}$ \& Annibale D'Ercole$^{3}$}
\affiliation{$^1$ Astrophysics, University of Oxford, Denys Wilkinson Building, Keble Road, Oxford OX1 3RH, U.K.
\break email: axp@astro.ox.ac.uk\\[\affilskip]
$^2$Dipartimento di Astronomia, Universita di Trieste, Via G.B. Tiepolo 11, 34100 Trieste, Italy \\[\affilskip]
$^3$INAF-Osservatorio Astronomico di Bologna, Via Ranzani 1, 40127 Bologna, Italy}

\pubyear{2007}
\volume{245}  
\pagerange{1--4}
\date{?? and in revised form ??}
\jname{Formation and Evolution of Galaxy Bulges}
\editors{M. Bureau, E. Athanasssoula \& B. Barbuy, eds.}
\begin{document}

\maketitle

\begin{abstract}
We present a new class of hydrodynamical models for the formation
of bulges (either massive elliptical galaxies or classical bulges in spirals) in which we implement detailed
prescriptions for the chemical evolution of H, He, O and Fe. 
Our results hint toward an outside-in
formation in the context of the
supernovae-driven wind scenario.
The build-up of the chemical properties of the stellar populations inhabiting
the galactic core is very fast. Therefore we predict a non significant evolution 
of both the mass-metallicity and the mass-[$\alpha$/Fe] relations after the first 0.5 - 1 Gyr.
In this framework we explain how the observed slopes, either positive
or negative, in the radial gradient of the mean stellar [$\alpha$/Fe],
and their apparent lack of any correlation with all the other
observables, can arise as a consequence of the interplay between
star formation and metal-enhanced internal gas flows.

\end{abstract}

\firstsection 
\section{Introduction}

Negative metallicity radial gradients in the stellar populations are a common
feature in spheroids (either elliptical and bulges, e.g. Carollo et al. 1993, Goudfroij et al. 1999 and Proctor et al. 2000, respectively).
Moreover bulges remarkably follow many other fundamental constraints for ellipticals
such as the mass-metallicity and the mass-[$\alpha$/Fe] relations, the only difference
being that they might be rejuvenated systems (Thomas \& Davies, 2006).
To date, only a handful of observational 
works inferred the gradients in the [$\alpha/Fe$] ratios both elliptical galaxies (Melhert et al. 2003, Annibali et al. 2006, 
Sanchez-Blazquez et al. 2007) and galaxy bulges (Jablonka et al. 2007).
These papers show that the slope in the [$\alpha/Fe$] gradient can be
either negative or positive, with a mean value close to zero, and that
it does not correlate with galactic properties (e.g. stellar mass for elliptical
and Hubble type of the host spiral for bulges). To date such facts
have not been properly addressed by any model of galaxy formation.

\section{The model}   

We adopted a one-dimensional hydrodynamical model which follows the
time evolution of the density of mass, momentum and
internal energy of a galaxy, under the assumption of spherical
symmetry.  Our main novelty is that
we follow the chemical evolution of several elements, namely H, He,
O and Fe. This set of elements is good enough to characterize our
simulated elliptical  galaxy from the chemical evolution point of view.  
We refer the reader to Pipino, D'Ercole \& Matteucci (2007, PDM) for a comprehensive
discussion of the adopted code and for the description of the models for massive spheroids.
Concerning the galaxy bulges, instead, we assume that they are stellar systems
with mass $\sim 2 \times 10^{10} M_{\odot}$ embedded in a $\sim 100$ times more massive Dark Matter halo.
We neglect the presence of a disc, which requires a much longer timescale to be built.
Several initial gas temperatures and density profiles have been tested.
In this paper we show the results for \emph{fiducial} model; this is the one, among all the
models we ran, which gives the best match of the Milky Way bulge.

\section{Results and discussion}

All the models undergo an outside-in formation as suggested by Pipino, Matteucci \& Chiappini (2006, PMC),
in the sense that star formation stops earlier in the outermost than
in the innermost regions, owing to the onset of a galactic wind
which proceeds from large radii (where the potential well is shallower) to the 
galactic core.

The mass-metallicity relation is satisfied, since our massive objects
have an average metallicity in the stars which is supersolar (see PDM for details), whereas
the simulated bulges have solar metallicity at most.
At the same time we also satisfy the relation between [$\alpha$/Fe] and mass
(see e.g. Thomas \& Davies, 2006), being the central region of our
massive spheroids more $\alpha$-enhanced that those of the bulges.
These relations are in place after 0.5 - 1 Gyr since the beginning
of the star formation.
It is importan to underline that our findings regarding the build-up 
of both the mass-metallicity and the [$\alpha$/Fe]-mass relations
are \emph{real predictions} and can be checked
by on-coming observational campaigns.

\subsection{Massive ellipticals}

We find [Fe/H] gradient slope in the range -0.5 -- -0.2 dex per decade
in radius and -0.3 dex per decade in radius for [Z/H], in
agreement with the observations (e.g. Annibali et al. 2007). 
Once transformed into
predictions on the line-strenght indices, these gradients in the
abundances lead to $d \rm Mg_2/log (R_{eff,*}/R_{core,*})\sim -0.06$ mag
per decade in radius, again in agreement with the typical mean values
measured for ellipticals by several authors and confirming the
PMC best model predictions.  The remarkable exception of some model
with a steeper gradient seems to go in the direction of a few massive
objects in the Ogando et al. (2005) sample.

\begin{figure}
 \includegraphics[width=7cm,height=5cm]{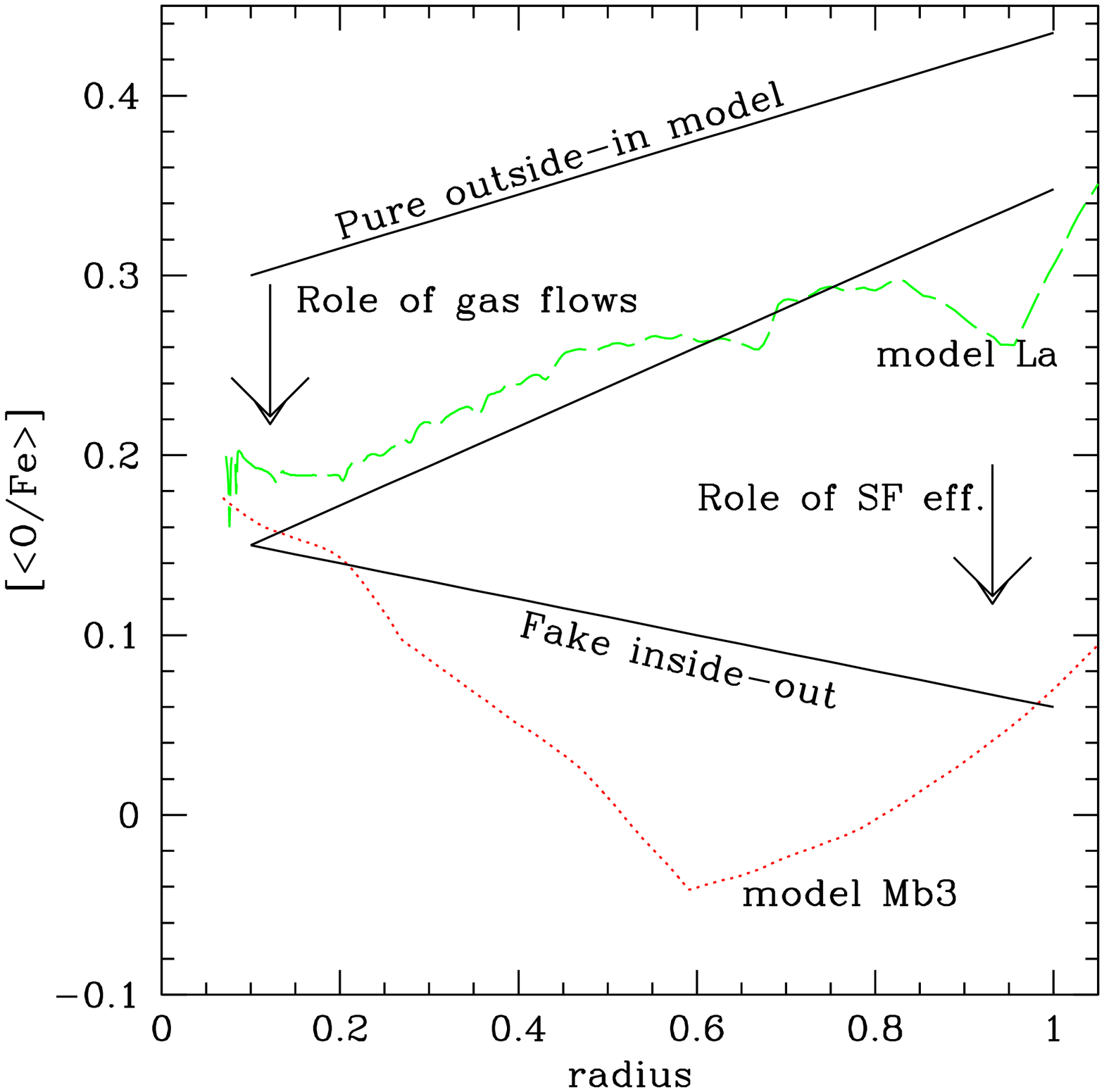}
\includegraphics[width=7cm,height=5cm]{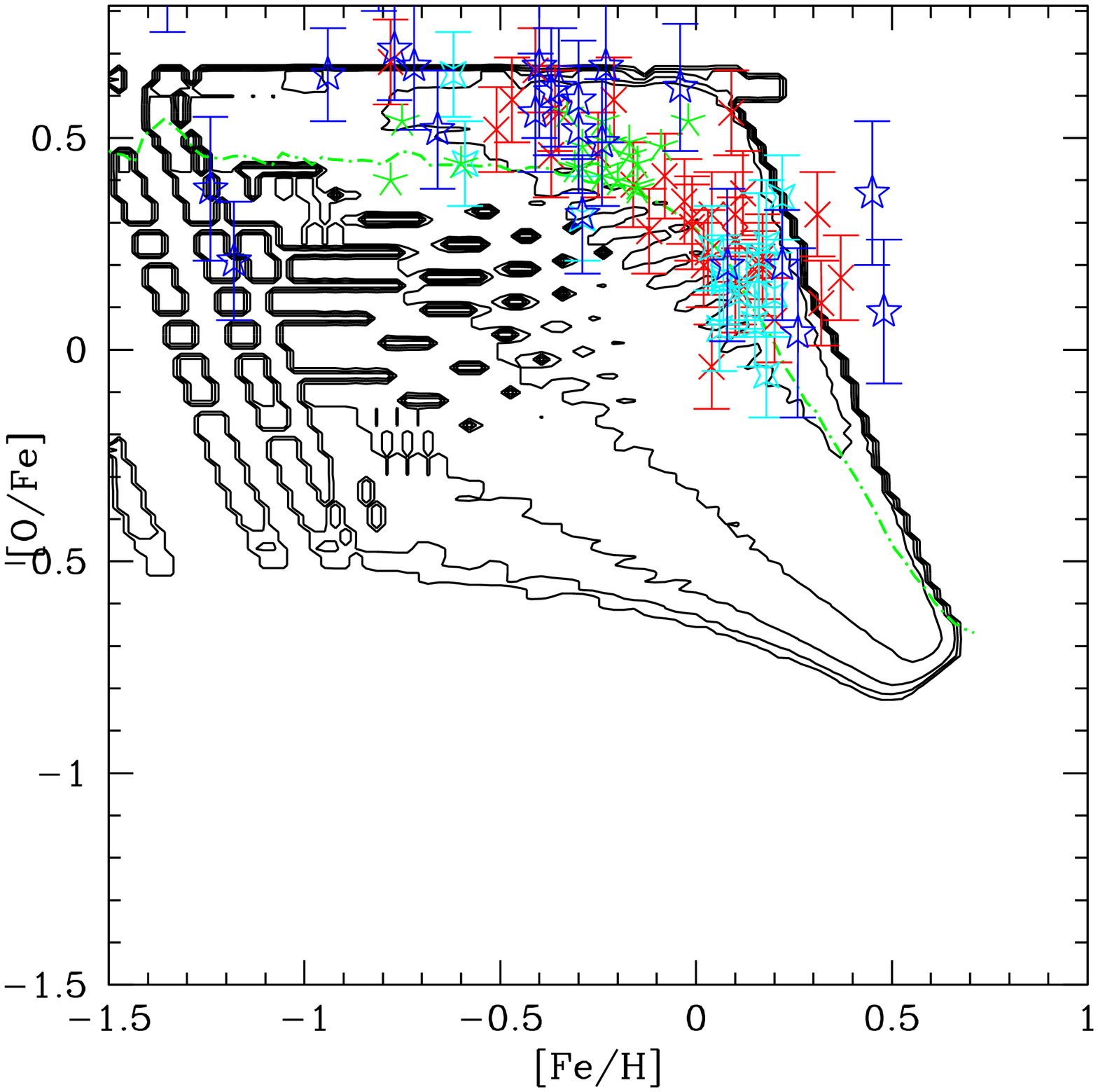} 
  \caption{\emph{Left panel} A sketch of the relative contribution of the gas flows strenght and 
the star formation efficiency $\nu $
to the creation of the final gradient for two particular cases from PDM.
\emph{Pure outside-in model:} hypothetical model with an outside-in formation, and $\nu $ constant with radius;
\emph{fake inside-out model:} hypothetical model with a strong variation of the star formation efficiency
with radius.
The abscissa is expressed in units of the effective radius.
\emph{Right panel} Contours: bidimensional metallicity distribution of stars as
functions of [Fe/H] and [O/Fe] for the MW bulge. 
Dashed line: [O/Fe] vs. [Fe/H] in the gas of model (mass-weighted values
on the gridpoints of each region).
Symbols: recent data compiled by Ballero et al. (2007).}
\label{fig_esplicativa}
\end{figure}

By analysing typical massive ellipticals, we find that all the models
that show values for their chemical properties, including the
[Fe/H] and the total metallicity gradients, within the observed
ranges, show a variety of gradient in the [$\alpha$/Fe] ratio, either
positive or negative, and one as no gradient at all.
We find that the predicted variety of the gradients in the [$\alpha$/Fe]  
ratio can be explained by physical processes, generally not taken into 
account in simple chemical evolution models, such as metal--enhanced 
radial flows coupled with different 
initial conditions.
Fig.~\ref{fig_esplicativa} (left panel) gives a sketch of the various mechanisms
at work (see PDM).

\subsection{Galaxy bulges}

Remarkably, all the results described in the previous section apply
to smaller objects (but embedded in much more massive haloes) such as
the galaxy bulges, although the gradient slopes are slightly smaller.
Investigation of a possible dependence between slope and mass
will be the subject of a forthcoming paper.
Therefore our models suggest the gradient in the [$\alpha$/Fe]
ratio to be related to the interplay between the velocity of the 
$\alpha$-enhanced radial flows, moving from the outer to the inner galactic 
regions, and the intensity and therefore duration of the SF formation process 
at any radius.
In other words, if the flow velocity is fast relative to the star formation, 
the stars still forming at inner radii have time to form out of 
$\alpha$-enhanced gas coming from the outermost regions, thus flattening and 
even reversing the sign of the [$\alpha$/Fe] gradient (fig.~\ref{fig_esplicativa}, left panel).

Moreover, the classical bulges allow us
a further tool to calibrate our models. In fact, we can compare our
model prediction to the properties of the \emph{resolved stellar population}
observed in the Milky Way bulge.
In fig.~\ref{fig_esplicativa} (right panel) we plot the distribution of stars formed out of gas with a
given chemical pattern (i.e. a given [$Fe/H$] and [$O/Fe$]) as
contours in the [$O/Fe$]-[$Fe/H$] plane. In particular, the contours
connect regions of the plane with the same mass fraction of stars. 
The dotted-dashed line is the average trend (which reflects the mean
composition of the gas), while the symbols are a compilation of
observations (see Ballero et al. 2007). 
We consider the stars
born in different points of the grid, which may have undergone
different chemical evolution histories. 
This explains why the lowest probability contour (i.e. the outermost)
is quite broad. 
In the single degenerate scenario, the minimum timescale for having a SNIa is 30 Myr, therefore
we can have stars forming in a cell in which [O/Fe] $<1$ even at [Fe/H]=-1.5, but this is very unlikely.
In fact, remarkably, the majority of the observed data points lie within the highest probability contour.
In practice our model can explain not only the average trend, but also the intrinsic scatter.
\emph{The minimum time is in general different for the timescale for SNIa to be effective}
which is tipically inferred from the position of the knee in the [O/Fe] vs [Fe/H] curve
(dashed line in fig.~\ref{fig_esplicativa}, right panel).

\begin{figure}
\includegraphics[width=7cm,height=4cm]{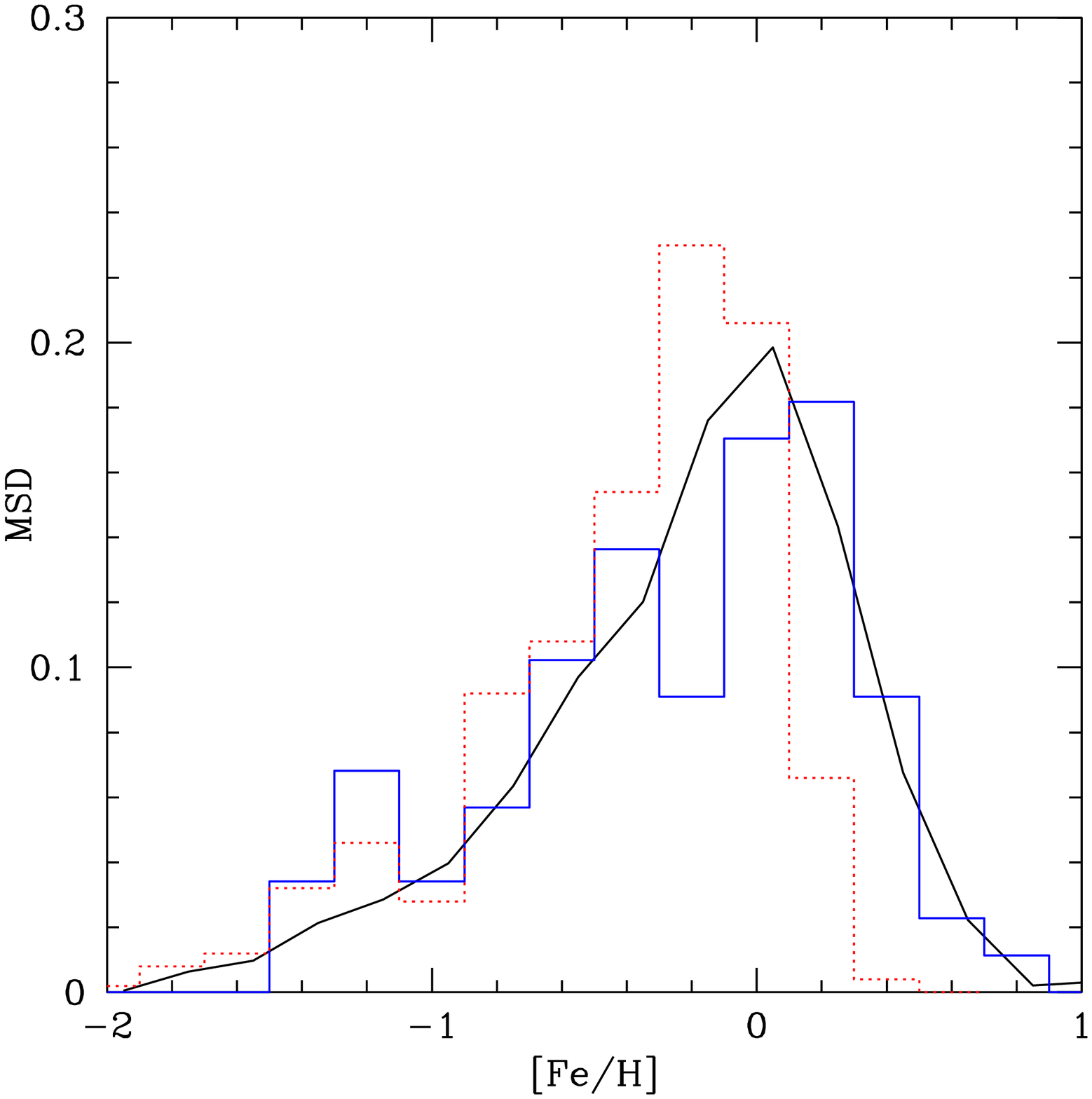} 
\includegraphics[width=7cm,height=4cm]{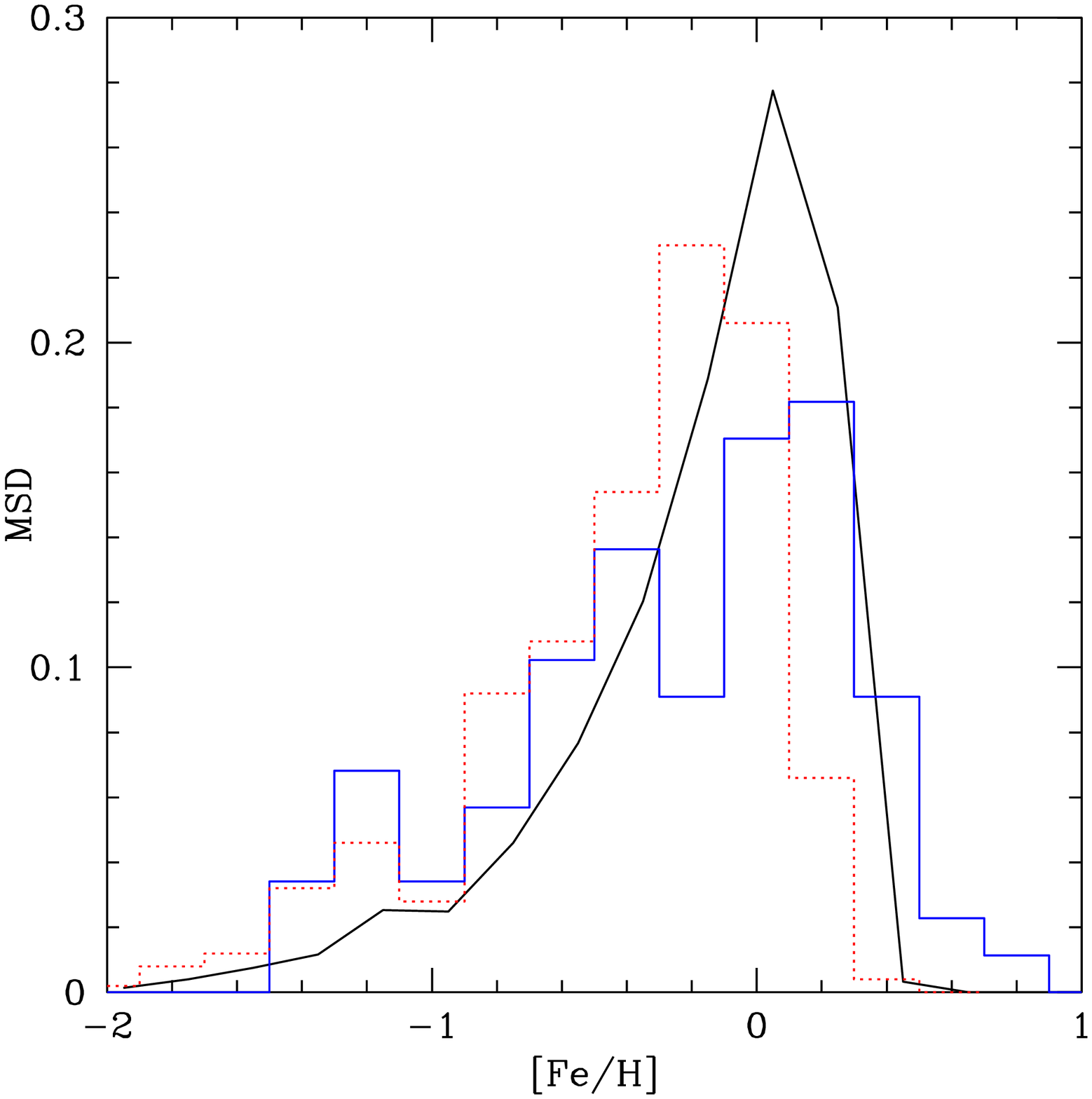} 
\includegraphics[width=7cm,height=4cm]{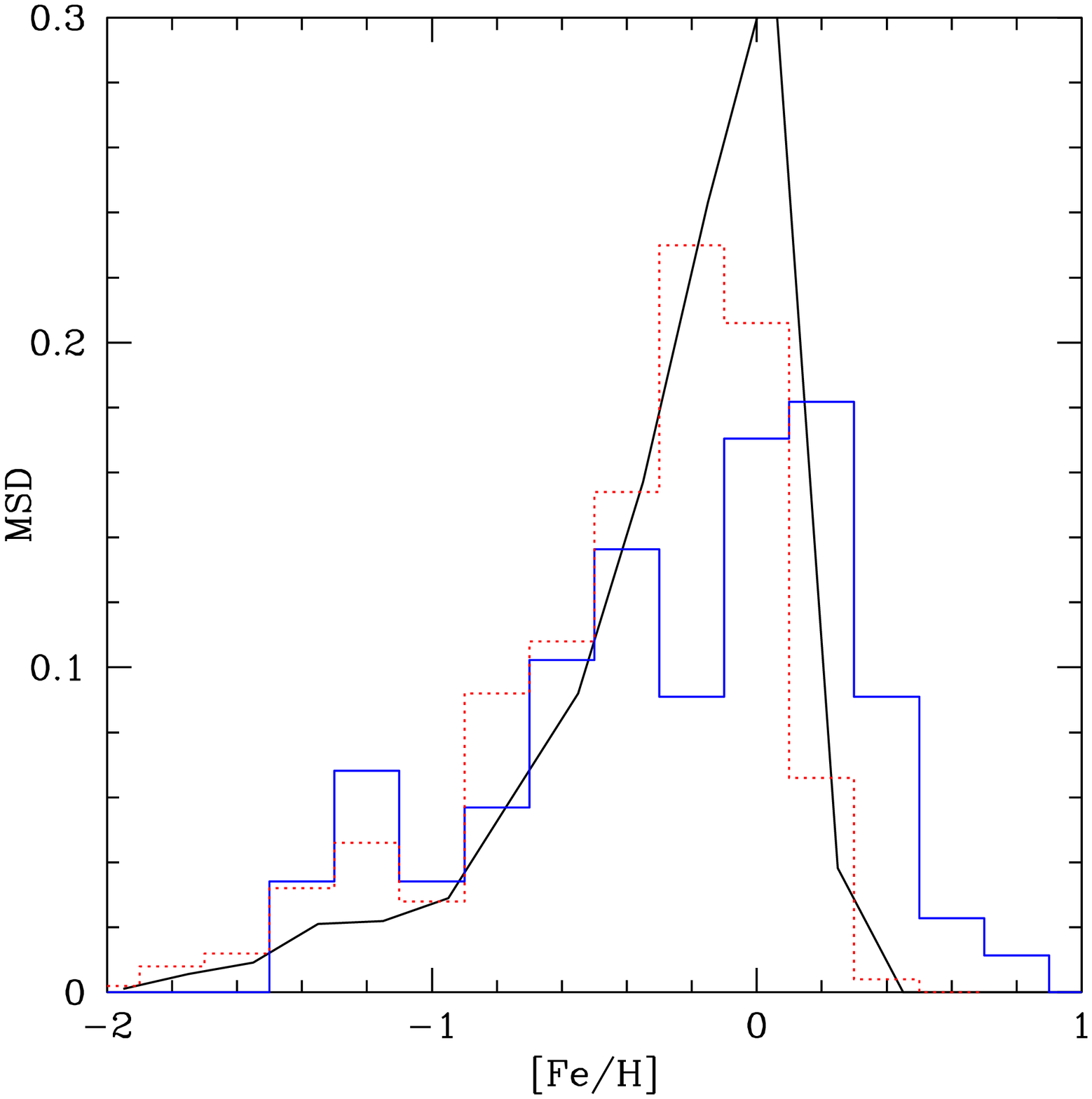} 
\includegraphics[width=7cm,height=4cm]{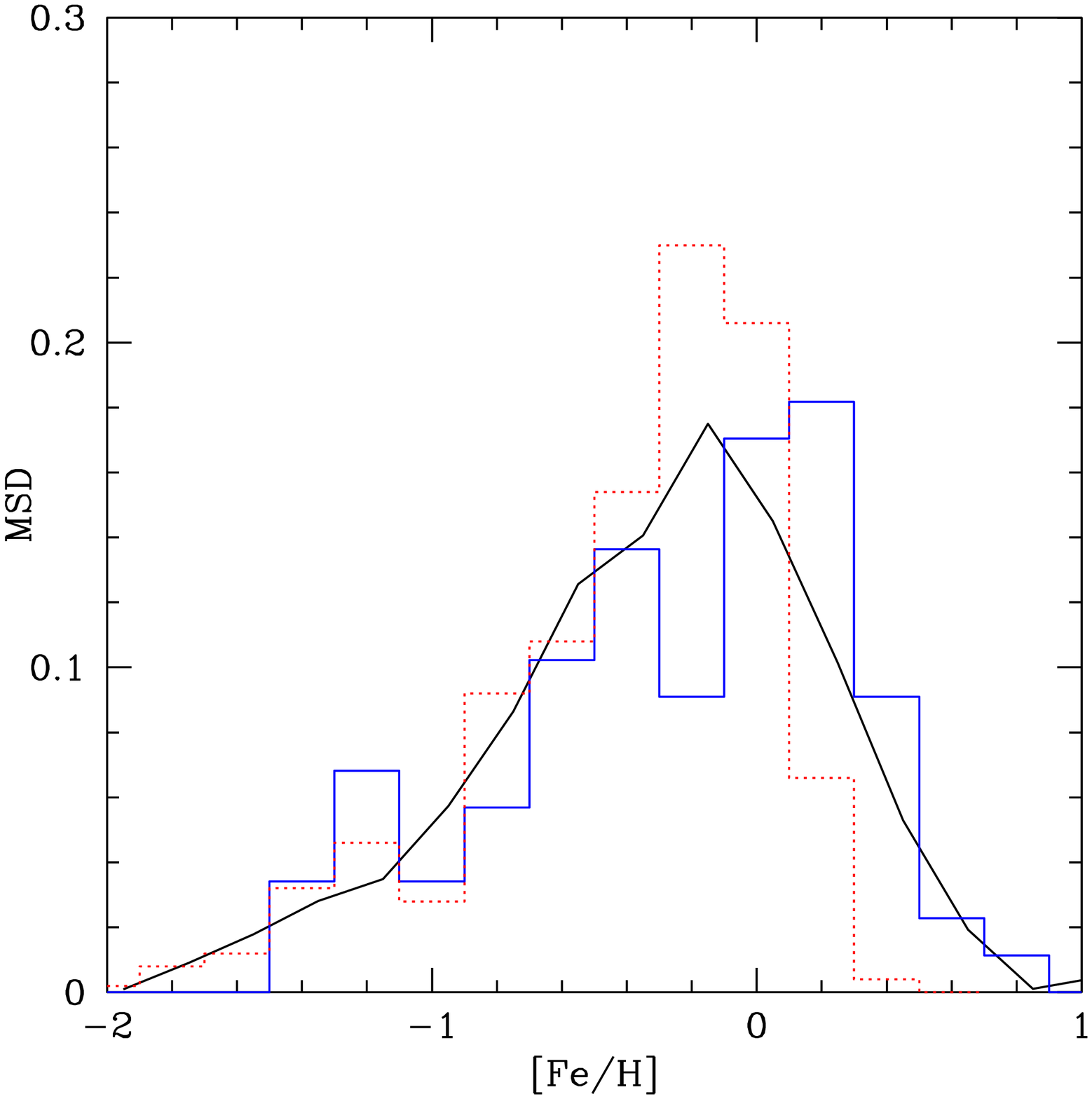} 
\caption{Metallicity distribution function for the stars in the MW bulge.
The histograms are the observed K-giants distributions by Fulbright et al. (2006, solid) and
Zoccali et al. (2003, dotted) compared to our fiducial model with different IMFs.
Upper-left: Salpeter (1955) over the whole mass range. Upper-right: x=0.33 for m$\le 1 M_{\odot}$, x=1 otherwise.
Lower-left: x=0.33 over the whole mass range. Lower-right: Kroupa (2001) IMF. }
\label{gdwarf}
\end{figure}

Our fiducial model features a Salpeter IMF, which successfully
reproduces the properties of massive spheroids.
In fig.~\ref{gdwarf} (upper-left panel) we compare the stellar metallicity
distribution predicted by such a model to two observed K-giants
distribution. In the other three panels of fig.~\ref{gdwarf}
we test other possible IMFs, mutivated by either observations
or theoretical efforts.
In agreement with Matteucci \& Brocato (1990) and Ballero et al. (2007),
we find that a Kroupa (2001) IMF poorly fits
the observations. On the other hand, an IMF with x=0.33 for m$\le 1 M_{\odot}$ and x=1 otherwise
represents another viable solution.
If the exponent x=0.33 is kept constant over the whole mass range predicts
too a narrow and peaked stellar metallicity distribution. Moreover the
typical [O/Fe] of the stars created with this last choice will be 
too high with respect to the observations.
Given the longer infall timescale, the differences between the IMFs
are less dramatic than those shown by Ballero et al. (2007).

Preliminary results shown by Lecureur et al (this book) seem to suggest
a radial variation of stellar metallicity distribution in qualitative
agreement with PMC and PDM finding. This topic will be investigated in the future.
Finally, we find that the bulge formation in fast in agreement with Matteucci \& Brocato (1990).
Star formation stops due to gas consumption rather than to galactic winds (Rich 1988;
Ballero et al. 2007).

\end{document}